\documentclass[preprint,12pt]{elsarticle}

\usepackage{graphicx}        
\usepackage{amsmath,amsfonts,amssymb}
\usepackage{mathptmx}      
\usepackage{subcaption}       
\usepackage{booktabs,tabularx} 
\usepackage{listings}         
\usepackage{xcolor}          
\usepackage[colorlinks=true,allcolors=blue]{hyperref}
\usepackage{xurl} 
\usepackage{hyperref}
\usepackage{float}
\usepackage{dblfloatfix}

\usepackage[
  top    = 20mm,
  bottom = 20mm,
  left   = 20mm,
  right  = 20mm
]{geometry}

\definecolor{dkgreen}{rgb}{0,0.6,0}
\definecolor{mauve}  {rgb}{0.58,0,0.82}
\journal{Magnetic Resonance Imaging}

\begin{document}
\begin{frontmatter}

\title{Fast Electromagnetic and RF Circuit Co-Simulation for Passive Resonator Field Calculation and Optimization in MRI}

\author[1]{Zhonghao Zhang \corref{equal}}
\author[2,3]{Ming Lu \corref{equal}}
\author[2,3]{Hao Liang \corref{equal}}
\author[2,3]{Zhongliang Zu}
\author[4]{Yi Gu}
\author[5]{Xiao Wang}
\author[1,6]{Yuankai Huo}
\author[1,2,3]{Xinqiang Yan \corref{corresponding}}

\cortext[equal]{These authors contributed equally to this work.}
\cortext[corresponding]{Corresponding author: xinqiang.yan@vumc.org}

\affiliation[1]{organization={Department of Electrical and Computer Engineering, Vanderbilt University}, city={Nashville}, state={TN}, country={USA}}
\affiliation[2]{organization={Vanderbilt University Institute of Imaging Science, Vanderbilt University Medical Center}, city={Nashville}, state={TN}, country={USA}}
\affiliation[3]{organization={Department of Radiology and Radiological Sciences, Vanderbilt University Medical Center}, city={Nashville}, state={TN}, country={USA}}
\affiliation[4]{organization={Department of Computer Science, Middle Tennessee State University}, city={Murfreesboro}, state={TN}, country={USA}}
\affiliation[5]{organization={Computational Science and Engineering Division, Oak Ridge National Laboratory}, city={Oak Ridge}, state={TN}, country={USA}}
\affiliation[6]{organization={Department of Computer Science, Vanderbilt University}, city={Nashville}, state={TN}, country={USA}}

\begin{abstract}
\noindent
\textbf{Purpose:} Passive resonators have been widely used in MRI to manipulate RF field distributions. However, optimizing these structures using full-wave electromagnetic (EM) simulations is computationally prohibitive, particularly for massive-element passive resonator arrays with many degrees of freedom.

\noindent
\textbf{Methods:} While the EM and RF circuit co-simulation method has previously been applied to RF coil design, this work presents, for the first time, a co-simulation framework tailored specifically for the analysis and optimization of passive resonators. The framework performs a single full-wave EM simulation in which the resonator's lumped components are replaced by ports, followed by circuit-level computations to evaluate arbitrary capacitor/inductor configurations. This allows integration with a genetic algorithm to rapidly optimize the resonator parameters to enhance \( B_1 \) fields in a targeted region of interest (ROI).

\noindent
\textbf{Results:} The proposed method was validated across three scenarios of increasing complexity: (1) a single-loop passive resonator on a spherical phantom, (2) a two-loop array on a cylindrical phantom, and (3) a two-loop array on a human head model. In all cases, the co-simulation results showed excellent agreement with full-wave EM simulations, with relative errors below 1\%. The genetic-algorithm-driven optimization, involving tens of thousands of capacitor combinations, completed in under 5 minutes—whereas equivalent full-wave EM sweeps would require an impractically long computation time.

\noindent
\textbf{Conclusion:} To the best of our knowledge, this work represents an early and systematic extension of the co-simulation methodology to passive resonator design, enabling fast, accurate, and scalable optimization. The approach significantly reduces computational burden while preserving full-wave accuracy, making it a powerful tool for passive RF structure development in MRI.
\end{abstract}


\end{frontmatter}

\section{Introduction}

Passive resonators, also known as inductively coupled or wireless resonators, have been widely used in MRI to manipulate the RF fields. By tuning the self-resonant frequency of a single-loop passive resonator, a secondary electromagnetic (EM) field can be generated that is either in phase or out of phase with the EM field of the primary coil. Passive resonators can also be configured as multi-unit arrays, offering greater flexibility for shaping the RF field. These arrays may be implemented in a decoupled manner, such as loop arrays with individually decoupled elements, or in a coupled design, such as ladder resonators or birdcage-type resonators.

From the perspective of using passive resonators to manipulate radiofrequency (RF) transmission, Wang et al. used a passive birdcage resonator to achieve localized RF transmission for extremity MRI without requiring wired local transmit coils \cite{wang2008inducitve}. Meanwhile, Wang et al. and Merkle et al. demonstrated that passive resonators can enhance peripheral $B^{+}$ fields, thereby improving transmit field ($B^{+}$) homogeneity within a volume coil at ultrahigh fields \cite{wang2008b_1, merkle2011transmit}. Beyond transmit field manipulation, one of the most important and widely adopted applications of passive resonators is to enhance the receive field ($B_1^{-}$) and thereby improve the signal-to-noise ratio (SNR). This concept dates back to the early days of the MRI field \cite{froncisz1986inductive, wright1991arrays}. It was attempted as fully detunable coils on human scanners around 2010s \cite{sahara2007development,zhu2012novel,zhu2012novel1, zhu2013wireless} and has recently regained significant attention \cite{yan2017improved,zhang2017sensitivity,jordan2019wireless, alipour2021improvement, brui2022volumetric, zhu2023helmholtz, zhu2024detunable, zhu2024detunable1, zhu2024detunable2, wu2024wireless,mo2025near}. It is important to note, however, that if such resonators are not detunable, they will inevitably alter the $B^{+}$ field during transmission. Besides the resonators (either loop or birdcage-type), the dipole antenna were also employed to enhance the $B_1^+$ and/or $B_1^-$ field in traveling wave MRI and MRgFUS \cite{yan2017improved,yan2023dark,lu2023investigating}.

The diverse and expanding applications of passive resonators in MRI have underscored the need for effective optimization strategies to maximize their performance. Full-wave EM simulation \cite{collins2016electromagnetics} has emerged as an indispensable tool in this process, providing detailed insights into the interactions between passive resonators and the primary RF coil, including how these interactions alter the original $B_{1}$ field. By enabling comprehensive analysis of $B^{+}$ or $B^{-}$ field distributions, EM simulation allows designers to optimize resonator configurations to enhance the transmit or receive performance. Consequently, simulation-driven design has become a critical component in the development pipeline of advanced passive resonator structures for MRI.

However, full-wave EM simulations are computationally demanding, especially when high mesh resolution and heterogeneous loading are required. As the number of design variables increases, the brute-force evaluation of numerous component combinations becomes impractical—particularly for arrays of passive resonators with many degrees of freedom. To overcome this limitation, we attempted to employ the EM and RF circuit co-simulation approach that enables fast and efficient simulation and optimization of passive resonator configurations in MRI, significantly reducing computational burden while maintaining high accuracy.

EM and RF circuit co-simulation was introduced to MRI in 2009 \cite{kozlov2009fast,zhang2009field} and has since been widely adopted for evaluating and optimizing RF coils \cite{lemdiasov2011numerical,yan2014optimization,shajan201416, beqiri2015comparison,yan2016simulation,golestanirad2017feasibility,li2021electromagnetic,zhang2021effect,zanovello2022cosimpy,lu2025magnetic,kazemivalipour2025simulated}. In this approach, lumped circuit elements (such as capacitors and inductors) are replaced with discrete ports during the time-consuming EM simulation. The frequency-dependent impedance or admittance characteristics at these ports are then extracted and incorporated into a separate circuit-level simulation. This enables rapid evaluation of various circuit configurations without the need to rerun computationally expensive full-wave EM simulations, thereby significantly accelerating the optimization process while preserving high accuracy.

Although passive resonators are not physically connected to the primary RF coils, their inductive coupling effectively forms a unified resonant system, with only a single real port associated with each primary coil. This EM interaction enables the passive resonators to be treated as an extension of the primary coil, making the use of co-simulation methods both appropriate and straightforward. However, it is important to note that conventional co-simulation frameworks—typically aimed at optimizing scattering (S-) parameters—may not be directly applicable to scenarios involving passive resonators. In these cases, the primary goal of co-simulation is to accurately characterize the EM field distribution, rather than to optimize network-level parameters.

We begin by reviewing the co-simulation methodology and then demonstrate its application to various types of passive resonators positioned inside a 3 Tesla body coil. We compare the simulation results and computation times between the co-simulation approach and conventional full-wave methods. Finally, we show how the co-simulation framework can be directly combined with a genetic algorithm to optimize the electromagnetic field distribution.

\section{Method}

\subsection{Review of EM and RF Circuit Co-Simulation Methods for MRI RF Coil}

EM and RF co-simulation can generally be categorized into two approaches based on the source of the circuit simulation and optimization tools: the use of commercially available software \cite{kozlov2009fast} or the development of in-house code \cite{zhang2009field,beqiri2015comparison,zanovello2022cosimpy}. Commercially available EM simulation software often integrates built-in tools for co-simulation. For instance, Ansys HFSS (Ansys, Inc., Canonsburg, PA, USA) and CST Studio Suite (Dassault Systèmes, Vélizy-Villacoublay, France) provide robust co-simulation functionalities, while Keysight Technologies (formerly Agilent, Santa Rosa, California, U.S) offers the Advanced Design System (ADS), a powerful tool for circuit simulation and optimization. Additionally, xFDTD (Remcom Inc., State College, PA, USA) is actively developing co-simulation capabilities to further enhance its software.

For conventional RF coil designs, where optimization objectives are typically centered around S-parameters—such as \( S_{xx} \) (return loss of each coil) and \( S_{xy} \) (transmission coefficients between coils)—these commercial tools are highly effective. Their intuitive, user-friendly interfaces have contributed to their widespread adoption in RF coil design workflows \cite{yan2014optimization,yan2016simulation}. However, it is worth noting that commercial tools are proprietary, not open-source, and often come with significant costs, which may limit accessibility for certain research groups or institutions. More critically, these tools demonstrate limitations in the optimization of passive resonators. Unlike conventional RF coils, which have physical ports that allow performance evaluation based on S-parameters, passive resonators lack such ports. The optimization objectives for passive resonators are instead focused on EM field characteristics, such as the \( B_1 \) field and/or \( E \) field distributions, rather than S-parameter metrics. Consequently, the conventional optimization approaches embedded within commercial software are incompatible with the unique requirements of passive resonator design.

In 2009, the Zhang et al. proposed a detailed methodology for co-simulation \cite{zhang2009field}, which was later adopted by Beqiri et al. \cite{beqiri2015comparison}, who subsequently provided a MATLAB implementation. More recently, in 2022, Zanovello et al. \cite{zanovello2022cosimpy} introduced a similar co-simulation mechanism, utilizing Python-based code and demonstrating its applicability for transmit RF coils. Both methods are built upon the same fundamental principles, with minor differences in the handling of matrix manipulation and transformation. In this work, we adapted the method from Zhang et al. \cite{zhang2009field}, and applied it for different passive resonators simulation and optimization.

\begin{figure}[htbp]
    \centering
    \includegraphics[width=1\linewidth]{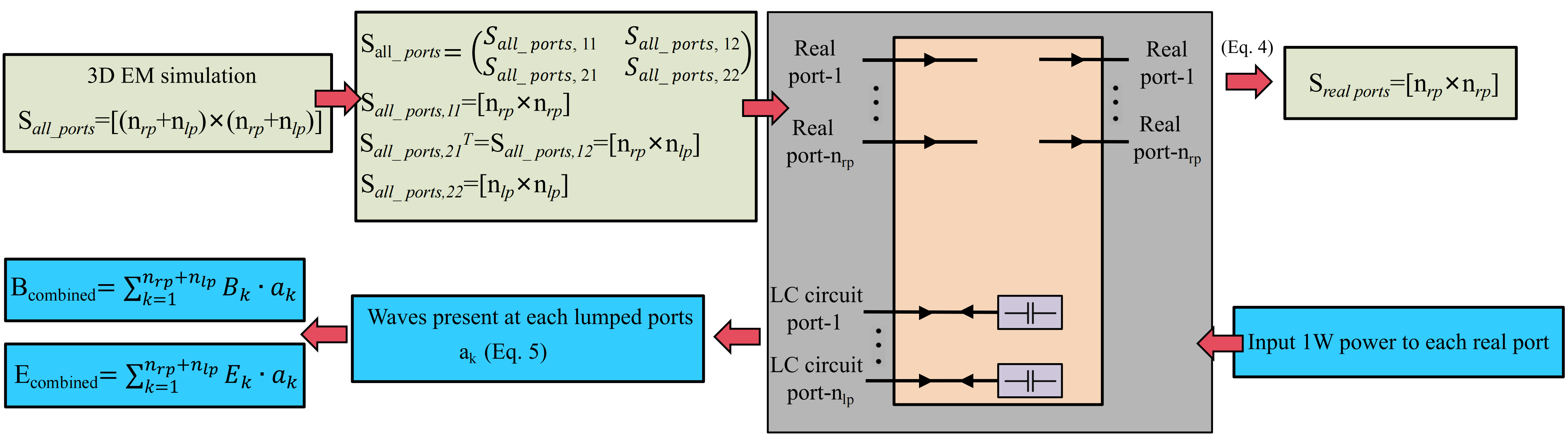}
    \caption{Schematic diagram of the co-simulation principle.}
    \label{fig:cosim Schematic}
\end{figure}

\subsection{Co-simulation for passive resonators}

A key advantage of the co-simulation method is that the behavior of lumped elements is handled within the circuit simulation environment. This eliminates the need for repeated full-wave EM simulations during the iterative process of adjusting component values in the passive resonators. Figure~\ref{fig:cosim Schematic} illustrates how the co-simulation approach enables accurate and efficient calculation of the \( B_1 \) field in the presence of passive resonators using only a single EM simulation.

First, we replace the lumped components in the passive resonators with ports and obtain the S-parameter matrix of all ports, denoted as \( S_{\text{all\_ports}} \). This matrix has dimensions $\left(n_{\text{rp}} + n_{\text{lp}}\right) \times \left(n_{\text{rp}} + n_{\text{lp}}\right)$, where \( n_{\text{rp}} \) is the number of real ports associated with the primary coils, and \( n_{\text{lp}} \) is the number of ports corresponding to the lumped components in the passive resonators, which are temporarily replaced with ports for co-simulation purposes.Importantly, the number of ports does not affect the number of required full-wave simulations, as all port interactions are captured within a single EM solution. Then, we subdivided \( S_{\text{all\_ports}} \) into terms that corresponded to primary coils and lumped components:

\begin{equation} \label{Stransf_def}
\left|
  \begin{array}{c}   
    b_{real\_ports}\\  
    b_{lumped\_ports}\\  
  \end{array}
\right|=
S_{all\_ports}
\cdot 
\left|
  \begin{array}{c}   
    a_{real\_ports}\\  
    a_{lumped\_ports}\\  
  \end{array}
\right|=
\left|
  \begin{array}{cc}   
    S_{all\_ports, 11} & S_{all\_ports, 12} \\  
    S_{all\_ports,21} & S_{all\_ports,22}\\  
  \end{array}
\right|  
\cdot 
\left|
  \begin{array}{c}   
    a_{real\_ports}\\  
    a_{lumped\_ports}\\  
  \end{array}
\right|,
\end{equation}
where \( a \) and \( b \) denote the incident and reflected waves, respectively. $S_{all\_ports,11}$ is a $n_\text{rp} \times n_\text{rp}$ matrix; $S_{all\_ports, 12}$ is $n_\text{rp} \times n_\text{lp}$; $S_{all\_ports, 21}$ is $n_\text{lp} \times n_\text{rp}$; $S_{all\_ports, 22}$ is $n_\text{lp} \times n_\text{lp}$. The vectors are in arbitrarily order that real ports indices 1 to \(\;n_\text{rp} \) and lumped ports indices \( (n_\text{rp}+1)\) to \((n_\text{rp}+n_\text{lp}) \).

Next, we compute the S-parameter matrix for the lumped components themselves, denoted as \( S_{\text{LC}} \). This matrix has dimension \( n_{\text{lp}} \times \ n_{\text{lp}}\) and can be derived from the impedance values of the lumped components using the following equation:

\begin{equation} \label{S_LC}
 S_{LC}=(\sqrt{y}\;Z\;\sqrt{y}+I)^{-1}(\sqrt{y}\;Z\;\sqrt{y}-I).  
\end{equation}

Here, \( I \) is the identity matrix of size \( n_\text{lp} \times n_\text{lp} \), and \( Z \) is the impedance matrix of the lumped components that were initially replaced by ports. The diagonal entries are defined as $Z_{nn} = R - \frac{j}{\omega C_n}$, where $C_n$ is the value of the $n$-th capacitor. The off-diagonal terms are zero. \( y \) is the characteristic admittance matrix whose diagonal elements are \( 1/z_0 \). The scalar \( z_0 \) represents the characteristic impedance of the ports, typically set to 50~\(\Omega\). The definitions of incident and reflected waves are reversed here because
the reflected wave from each port becomes the incident wave on the lumped components:

\begin{equation} \label{S_LC_Relation}
 a_{lumped\_ports}=S_{LC}\,\cdot\,b_{lumped\_ports}.  
\end{equation}

Based on the the relations above, we can calculate the \( n_{\text{rp}} \times  n_{\text{rp}}\) S-Matrix. The details of the derivation can be found in Appendix~I.

\begin{equation} \label{Real_port_S}
S_{{real\_ports}}  = S_{{all\_ports, 11}} + S_{{all\_ports, 12}}\cdot S_{{LC}}(I-S_{all\_ports,22} \cdot S_{{LC}})^{-1} S_{{all\_ports, 21}} 
\end{equation}

After obtaining \( S_{\text{real\_ports}} \), we can compute the complex excitation weights at each real port and each LC port as shown below. It was generated by calculating the
 waves present at each lumped ports when real port was set to 1W input power. The details of the derivation can be found in Appendix~II. 

\begin{equation} \label{B_Weight}
 a_{lumped\_ports}=S_{LC}\,\cdot\,(I-S_{all\_ports,22}\cdot S_{LC})^{-1}\cdot S_{all\_ports,21}\cdot a_{real\_ports}. 
\end{equation}

Once, \( a_{\text{lumped\_ports}} \) was determined, we can reconstruct the total fields (either \( B_1 \) or electric fields) through superposition. Note that in the initial full-port simulation, each port (including both real ports and LC ports) was excited with 1-watt input power while all other ports were terminated with 50~\(\Omega\). Therefore, the resulting \( B_1 \) field corresponds to 1-watt input power.

\begin{equation}
B_{\text{combined}} = \sum_{k=1}^{\,{n_{\text{rp}}+n_{\text{lp}}}} B_k \cdot a_k
\quad \text{and} \quad
E_{\text{combined}} = \sum_{k=1}^{\,{n_{\text{rp}}+n_{\text{lp}}}} E_k \cdot a_k
\end{equation}

\subsection{Validation with Passive Resonators Inside 3~T Body Coil, and Maximization of \( B_1 \) in Nearby Region}

\begin{figure}[htbp]
    
    \includegraphics[width=1\linewidth]{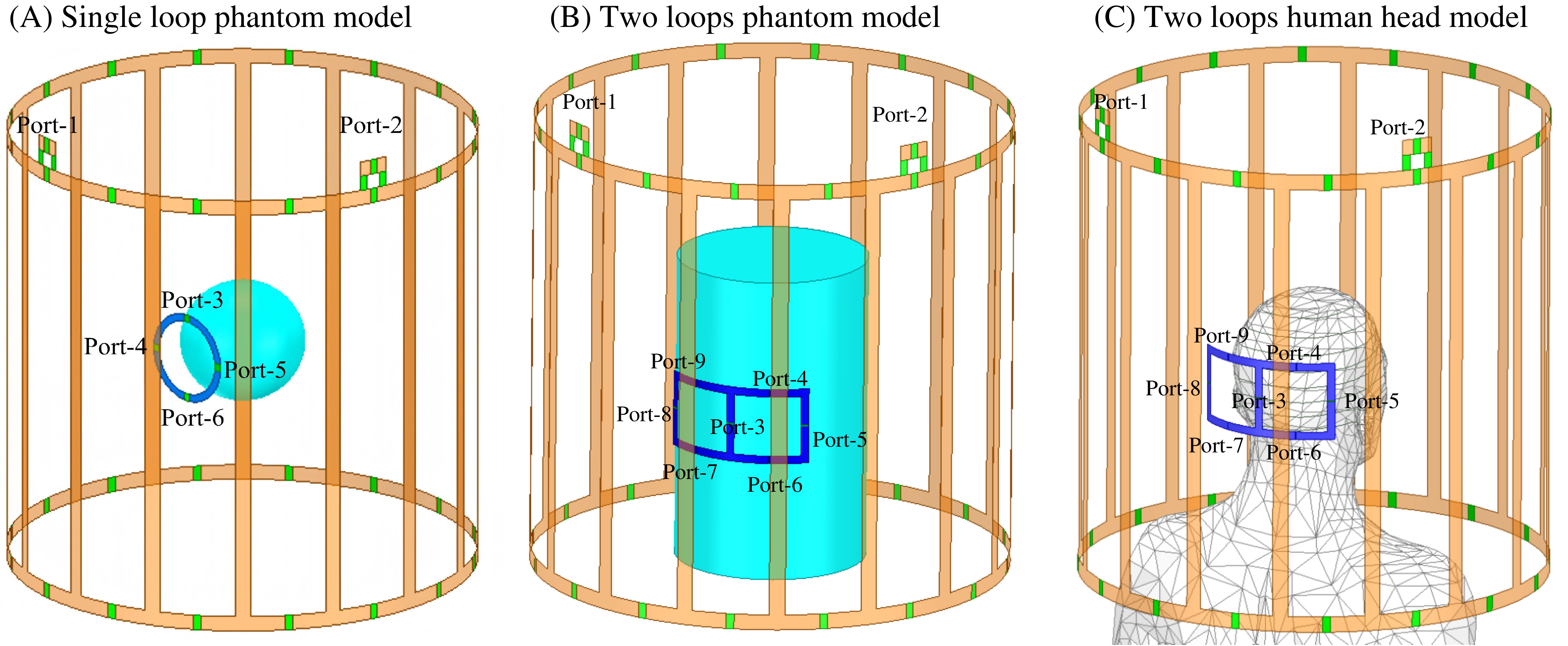}
    \caption{Model [A] shows 6 ports (where ports 1-2 are for the birdcage coil, and ports 3-6 are for the ports on wireless coils) used in the ultrafast simulation. Ports on wireless coils will be replaced by capacitors in 2-port validation model. Model [B] and Model [C] shows 9 ports(where ports 1-2 are for the birdcage coil as well, port 3 is for decoupling, and ports 4-9 are for the ports on wireless coils). Similar to previous model, ports 4-9 will be replaced by capacitors in validation stage.}
    \label{fig:simulationModel}
\end{figure}

To validate the accuracy and significantly improved speed of the co-simulation approach, we applied the proposed two-stage workflow to multiple passive resonator simulation scenarios inside a 3~T body coil and compared the results with those obtained from full-wave EM simulations. The body coil was modeled as a 16-rung high-pass birdcage coil with a diameter of 60~cm and a length of 60~cm, as shown in Figure~\ref{fig:simulationModel}. The conductors were modeled as copper (a finite conductivity of \( 5.8 \times 10^7~\text{S/m} \)) with a width of 2~cm. The two ports of the birdcage coil (port I and port Q) were tuned to 128~MHz , matched to 50~\(\Omega\) (return loss less than -20 dB), and decoupled with isolation greater than 20~dB in the absence of passive resonators, across different loading conditions.

\subsubsection{Validation of Simulation Accuracy for a Single-Loop Passive Resonator on a Spherical Phantom}

This approach was first validated using a single loop passive resonator placed inside the 3~T body coil, as shown in Figure~\ref{fig:simulationModel}A. The passive resonator loop had a diameter of 12~cm and included four evenly spaced tuning capacitors of the same value (\( C_{t} \)). The passive loop resonator was modeled as copper. A spherical phantom with a diameter of 16~cm was positioned 2~cm below the resonator and assigned a relative permittivity (\( \varepsilon_r \))  of 81 and a conductivity (\( \sigma \)) of 0.5~S/m.

In the first stage of the co-simulation workflow, the four capacitors in the resonator were temporarily replaced by ports. A \( 6 \times 6 \) S-parameter matrix, including two real ports from the birdcage coil and four ports in the resonator, was extracted from a full-wave simulation in ANSYS HFSS, along with the magnetic field distributions corresponding to each port excitation. In the second stage, the S-parameters of the real ports (1 and 2) and the resulting \( B_1 \) fields under different values of \( C_{t} \) (0~pF, 10~pF, 20~pF, 25~pF, and 100~pF) at ports 3--6 were computed using Eqs.~1--6). To validate the accuracy of the co-simulation method, full-wave simulations with identical capacitor values were performed for comparison, with identical meshing setting.

\subsubsection{Maximization of \( B_1 \) Field from a Two-Loop Passive Resonator Array in a Nearby Region of Interest}

The co-simulation method was then applied to maximize the local \( B_1 \) of a two-loop passive resonator array, as shown in Figures~\ref{fig:simulationModel}B and C. Each passive resonator measured 10~cm~\(\times\)~10~cm and was modeled as copper and a trace width of 1~cm. The two resonators together incorporated six distributed capacitors (\( C_{t} \)) and one shared capacitor (\( C_d \)). Note that \( C_d \) was used to adjust the coupling or decoupling between the two resonators, while \( C_{t} \)) were primarily responsible for frequency tuning. In the first stage of the co-simulation workflow, a \( 9 \times 9 \) S-parameter matrix was generated and exported. Two types of loading were investigated. First, a cylindrical phantom with a diameter of 24~cm and a length of 40~cm (\( \varepsilon_r \) = 78 and \( \sigma \) of 0.6~S/m) was used (Figure~\ref{fig:simulationModel}B). In addition to the cylindrical phantom, we also conducted simulations using a human head model provided by ANSYS (Figure~\ref{fig:simulationModel}C). For the phantom model, the two-loop wireless resonator array was placed 2.5 cm beneath the phantom, while for the head model, it was placed 4 cm beneath the head.

For local passive resonators, a common objective is to maximize the \( B_1 \) field magnitude in a specific nearby region. To demonstrate the utility of the proposed co-simulation framework in optimizing the resonators' component values for this purpose, we formulated the following cost function and applied a genetic algorithm (GA) for optimization:

\begin{equation}
\max\limits_{C_{\mathrm{t}},\,C_{\mathrm{d}}}
\left\lVert
  B_{\mathrm{1, ROI}}
\right\rVert,
\label{eq:cost_function}
\end{equation}
\noindent

where \( B_{\text{1,ROI}} \) denotes the \( B_1 \) field magnitude within the region of interest (ROI). In this work, the ROI is defined as the area immediately surrounding the passive resonators. The genetic algorithm (GA) was used to optimize two variables, \( [C_{t},\,C_{d}] \), within the bounds of 0.001--50~pF. The optimization score in our formulation typically ranged from 0 to 2. The termination criterion was set to a function tolerance of \( 10^{-6} \). A maximum of 150 generations was further imposed as a safeguard. At each iteration, candidate values of \( [C_{t},\,C_{d}] \) were used to update \( S_{\text{LC}} \), representing the circuit behavior of the lumped components. The resulting \( B_1 \) field was then computed via co-simulation, and the cost function (Eq.~\ref{eq:cost_function}) was evaluated to determine the performance score for that specific capacitor combination.

\section{Results}

\subsection{Single-loop passive resonator validation}

\begin{figure}[H]
    \centering
    \includegraphics[width=1\linewidth]{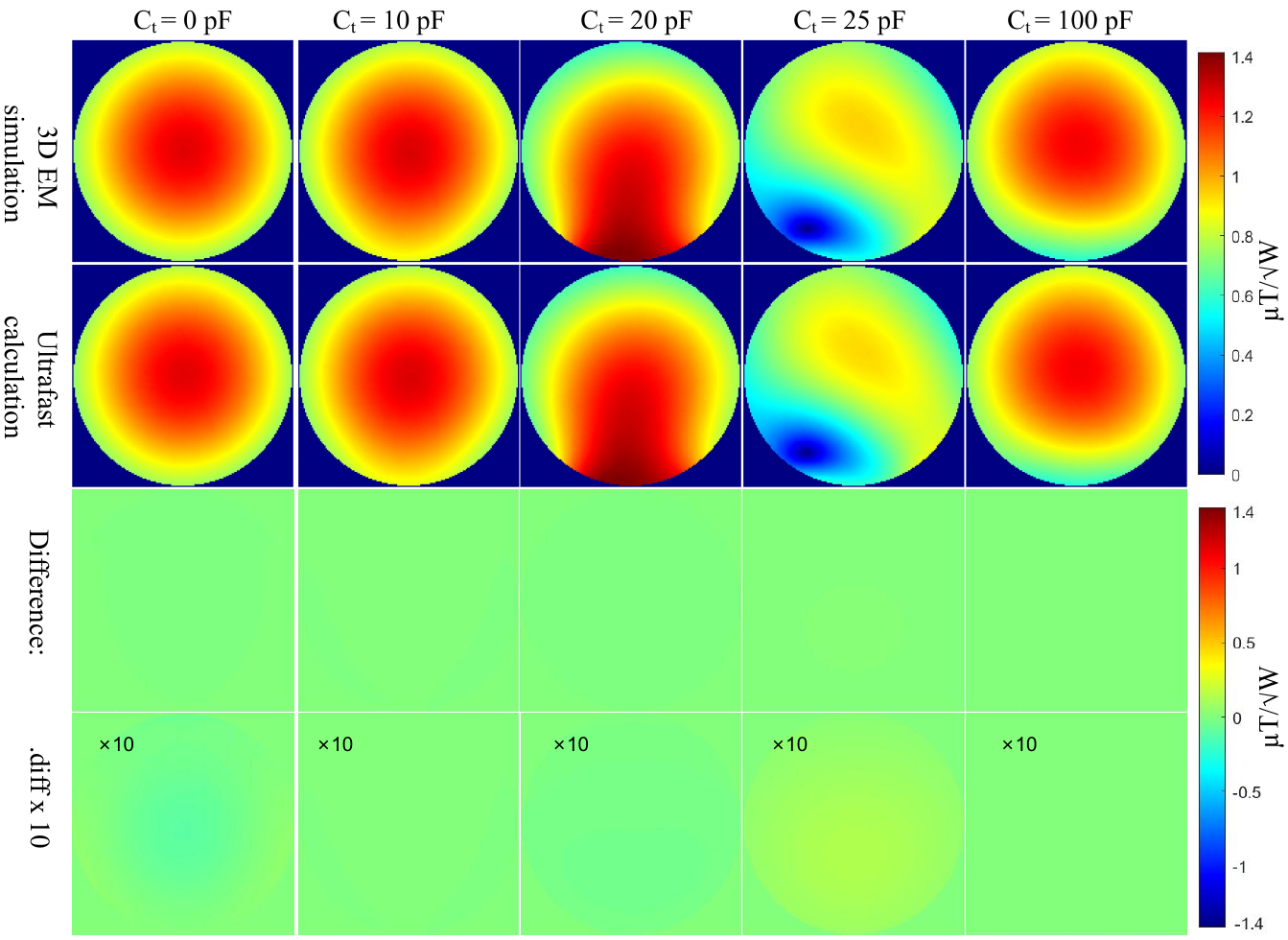}
    \caption{$B_{1}$
 fields from EM simulation and ultrafast simulation across varying capacitance values. The
maximum relative percentage difference is only 0.9\%, showing close alignment and confirming the reliability
of the ultrafast simulation for $B_{1}$
 field prediction.}
    \label{fig:B1_singleLoopl}
\end{figure}

Figure~\ref{fig:B1_singleLoopl} compares the \( B_1 \) field of the birdcage coil in the presence of a single-loop passive resonator, using full-wave EM simulation (top row) and the proposed co-simulation approach (bottom row). The tuning capacitor \( C_t \) in the passive resonator was used to control the resonant behavior of wireless coil and varied over a wide range, from an extremely small value (approximately 0~pF) to a large value of 100~pF. When \( C_t = 0~\text{pF} \), the resonator behaves as an open circuit and does not influence the original body coil field, serving as the baseline scenario. Based on analytical calculations, the required capacitance for self-resonance at 128~MHz is approximately 22.5~pF. As expected, a slightly smaller \( C_t \) enhances the nearby \( B_1 \) field, whereas a slightly larger \( C_t \) reduces it. This behavior arises from the change in the resonant frequency caused by impedance tuning. It is also consistent with resonator physics: a slightly smaller \( C_t \) shifts the resonance above the Larmor frequency, making the resonator act as an inductive enhancer; in contrast, a slightly larger \( C_t \) shifts the resonance below the Larmor frequency, causing the resonator to behave more like a reflector. When \( C_t \) deviates substantially from the self-resonant value (e.g., 10~pF or 100~pF), the passive resonator has minimal impact on the body coil’s field. From an RF transmission perspective, the passive resonator acts as a secondary source that generates additional electromagnetic fields and interacts with the primary field, thereby modifying the magnetic field sensitivity.

The third row of Figure~\ref{fig:B1_singleLoopl} shows the difference in the \( B_1 \) field between the full-wave EM simulation (top row) and the  co-simulation approach (bottom row). The fourth row presents the same difference scaled by a factor of 10 for improved visibility. Across all tested \( C_t \) values in the passive resonator, the co-simulation method demonstrates excellent agreement with the full-wave simulation, yielding an average and maximum relative percentage difference of only 0.24\% and 0.9\%, respectively.

\begin{figure*}[htbp]
    \centering
    \includegraphics[width=1\linewidth]{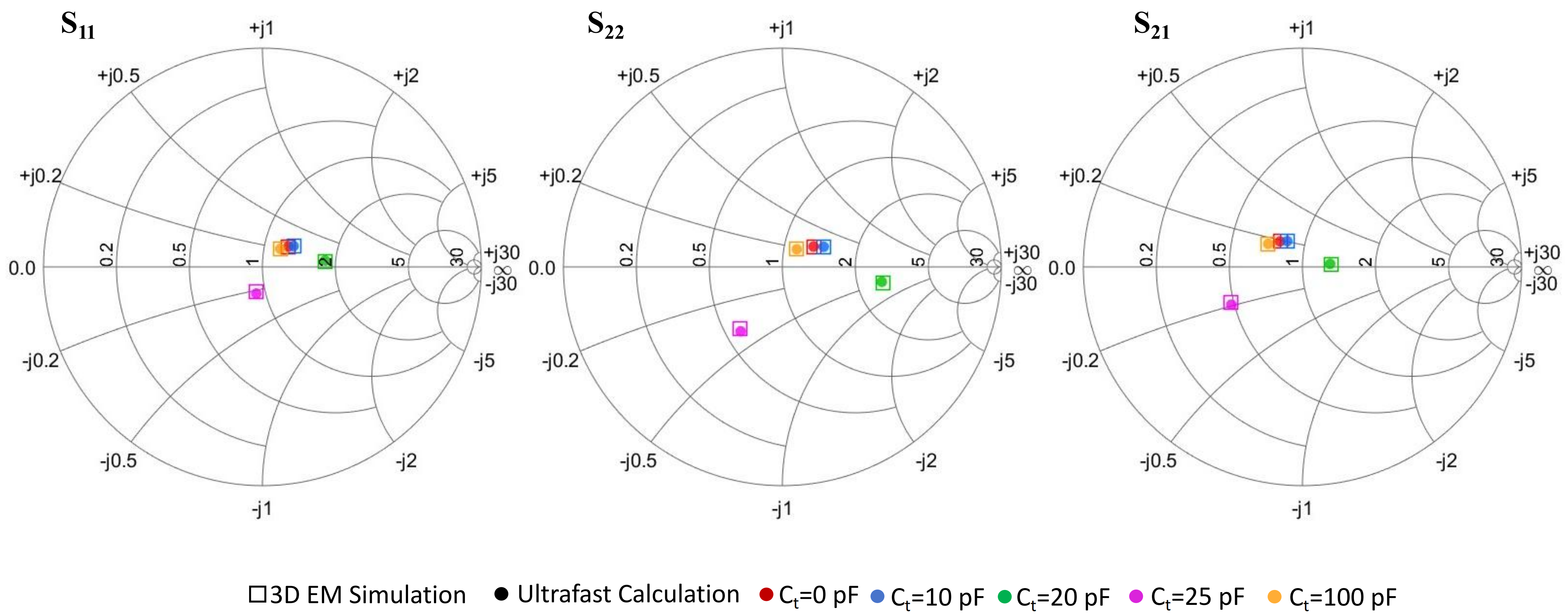}
\caption{Comparison of S-parameters on the Smith Chart for the primary body coil in the presence of various passive resonators. Different colors are shown as different \( C_{t} \) values. Results from full-wave EM simulations are shown as outlined boxes, while results from the co-simulation method are shown as dots. Excellent agreement is observed between the two methods for \( S_{11} \), \( S_{12} \), and \( S_{22} \) across different tuning capacitor values (\( C_t \)).}

    \label{fig:Spara_singleLoop}
\end{figure*}

In addition to the \( B_1 \) field comparison, Figure~\ref{fig:Spara_singleLoop} also compares the S-parameters of the body coil for different \( C_t \) values, obtained using both full-wave EM simulation and the proposed co-simulation method. Consistent with the field results, the S-parameters from both approaches show excellent agreement for each \( C_t \), confirming the accuracy and reliability of the co-simulation method. Notably, while the full-wave simulation for each configuration required up to 5 hours of computation time, the co-simulation produced results in approximately 10 milliseconds, once the $6 \times 6$ S-parameter matrix was determined. This dramatic reduction in computation time enables efficient optimization of passive resonator component values, which may require evaluating millions of combinations during the design process.

\subsection{Two-loop passive resonator array optimization}

\begin{figure*}[!t]
    \centering
    \includegraphics[width=1\linewidth]{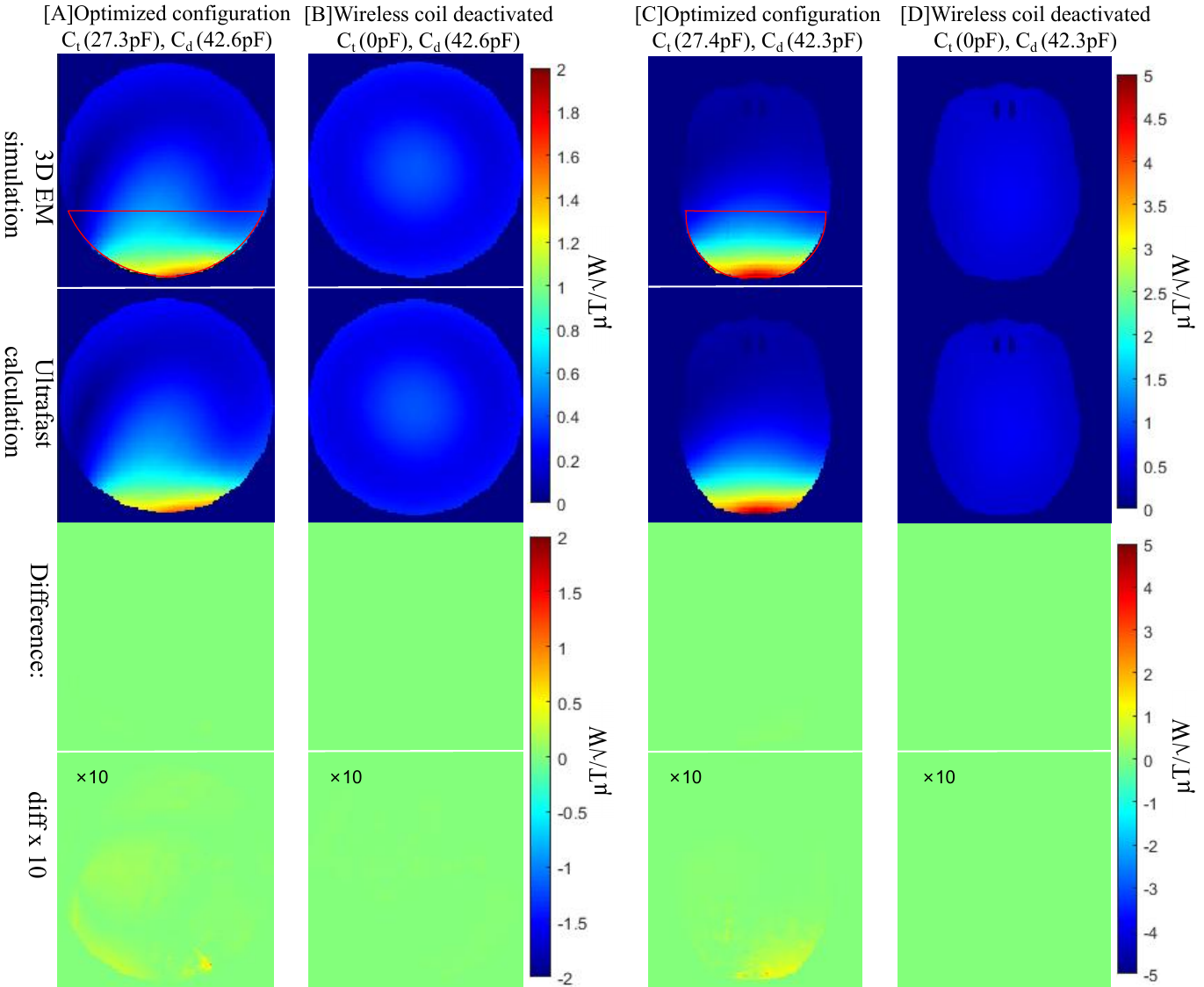}
    \caption{Comparison of full‑wave EM simulation and ultra-fast calculation results for the optimal passive resonator configuration versus the resonator‑off case in the spherical phantom and human head models. The highlighted regions of interest (ROI) show where we calculated the relative error for the optimal configuration. Passive resonators create a near‑zero field region that would distort the global relative error, so it is excluded; this excluded region exhibits the largest absolute errors. Average ROIs' \( B_1^+ \) field for the phantom and human head scenarios improved approximately \textbf{2.5}-fold and \textbf{3.7}-fold respectively, demonstrating that the ultrafast approach accurately reproduces full‑wave EM results while enhances \( B_1^+ \) field.}
    \label{fig:B1_GA_results}
\end{figure*}

Figure~\ref{fig:B1_GA_results} shows the \( B_1^+ \) field distributions in both the cylindrical phantom and human head model, using the optimized capacitor values obtained from the GA tuner in MATLAB. The GA was used to maximize the \( B_1^+ \) field within the defined region of interest (ROI). For the phantom case, the optimal capacitor values were \( C_t = 27.3~\text{pF} \) and \( C_d = 42.6~\text{pF} \); for the human head model, they slightly shifted to \( C_t = 27.4~\text{pF} \) and \( C_d = 42.3~\text{pF} \). Compared to the baseline case with the body coil alone (see the second and fourth columns in Figure~\ref{fig:B1_GA_results}), the inclusion of optimized passive resonators led to an improvement of approximately \textbf{2.7}-fold and \textbf{3.7}-fold in the local \( B_1^+ \) field for the phantom and human head scenarios, respectively.

We further validated the co-simulation's accuracy by comparing the results with full-wave EM simulations under the same capacitor configurations. Consistent with earlier findings from the single-loop resonator case, the optimized \( B_1^+ \) fields from co-simulation differed by only 0.5\% on average compared to the full-wave simulation within the ROI. Notably, the entire GA-driven co-simulation process, which involved tens of thousands of capacitor evaluations, completed in under 5 minutes. In contrast, using full-wave EM simulation for the same number of evaluations would require over 6 years of continuous computation time (assuming 5 hours per simulation), making such optimization impractical without the proposed fast co-simulation framework.

\section{Discussion}

\subsection{Summary}

In summary, this work validates an EM and RF circuit co-simulation framework specifically tailored for passive resonator design in MRI. To our knowledge, this is the first application of such a co-simulation method to passive resonators, although similar approaches have been widely used in RF coil design. By decoupling the time-consuming EM simulation from the ultra-fast lumped element circuit evaluation, the proposed method enables rapid and accurate assessment of EM field distributions across large parameter spaces. Extensive validation using both phantom and human head models demonstrated sub-1\% error compared to full-wave simulations, while reducing the runtime of a single configuration from several hours to just milliseconds. This framework offers a practical and scalable solution for optimizing passive resonators in MRI systems. When combined with a genetic algorithm or other optimization methods, it allows efficient tuning of resonator configurations to manipulate EM fields, such as maximizing \( B_1 \) enhancement.

\subsection{Algorithm}

In this study, we applied a genetic algorithm (GA) to optimize the capacitor values \( C_t \) and \( C_d \) in order to maximize the local \( B_1 \) field. This approach was sufficient for our application, as the optimization problem involved only two variables and a relatively straightforward objective function—maximizing \( |B_1| \) within a defined region of interest. However, in more complex passive resonator systems—such as large arrays with many units, or applications requiring field enhancement across multiple regions of interest or with specific spatial shaping constraints—the number of optimization variables increases significantly, and the cost function may become highly nonlinear, multi-modal, or multi-objective. In such cases, a standard GA may not offer adequate convergence speed or accuracy. To address these challenges, more advanced optimization strategies may be required. These could include hybrid approaches (e.g., GA combined with gradient-based refinements), Bayesian optimization, or deep learning–based surrogate models that learn the mapping between resonator configurations and resulting field distributions. Integrating such methods with the proposed co-simulation framework would enable scalable and efficient optimization of more complex passive resonator designs.

\subsection{Extension to Other Applications}

Although this study was demonstrated using single-loop and two-loop passive resonators, the proposed co-simulation framework is readily extendable to more complex geometries and larger multi-element arrays. For instance, the method can be applied to ladder-type, birdcage-inspired, or metamaterial-based resonators, where the number of elements and structural complexity are significantly greater. The underlying principle—replacing lumped components in passive resonators with ports during the EM simulation and evaluating different configurations through fast circuit-level modeling—remains applicable regardless of the array’s size or complexity.

\subsection{Limitations of This Work}

It should be noted that the proposed co-simulation method is specifically designed for passive resonators constructed using lumped elements, primarily discrete capacitors and inductors. However, the current approach does not directly support the optimization of geometric parameters—such as the shape, size, trace width, or spatial layout of the resonator elements—which significantly influence the electromagnetic behavior. Modifying these structural features would require repeated full-wave EM simulations to accurately capture their impact on field distributions. Therefore, while our method greatly accelerates the optimization and tuning of component values in a fixed resonator layout, it does not eliminate the need for conventional full-wave simulations when geometric changes are involved. 

\section{Conclusion}

We presented a fast and accurate EM and RF circuit co-simulation framework tailored for the design and optimization of passive resonators in MRI. While co-simulation has been widely used for RF coil design, this is the first study to extend it to passive wireless resonators. By decoupling the full-wave EM simulation from lumped-element tuning, the framework enables rapid evaluation of component configurations. Validation across single- and two-loop resonator setups, including a realistic human head model, demonstrated excellent agreement with full-wave simulations (sub-1\% error) while reducing simulation time from hours to milliseconds. When combined with a genetic algorithm, the framework efficiently optimized capacitor values to enhance local \( B_1 \) fields, completing tens of thousands of evaluations in under 5 minutes. This approach is readily extendable to more complex resonator geometries and larger arrays and offers a practical tool for accelerating passive RF structure design in both transmit and receive MRI applications.

\section{Acknowledgments}
This work was supported by the National Institutes of Health under grant numbers R01 EB031078, R21 EB029639, R03 EB034366, R21 EB037763 and S10 OD030389. The content is solely the responsibility of the authors and does not necessarily represent the official views of the National Institutes of Health.

\section{Declaration of generative AI and AI-assisted technologies in the writing process}

During the preparation of this work, the author(s) used ChatGPT4/ChatGPT4o in order to check for writing typos. After using this tool, the author(s) reviewed and edited the content as needed and take(s) full responsibility for the content of the published article.

\bibliographystyle{unsrt}
\bibliography{refs}

@article{wang2008b_1,
  title={$ B\_1 $ homogenization in MRI by multilayer coupled coils},
  author={Wang, Shumin and Murphy-Boesch, Joseph and Merkle, Hellmut and Koretsky, Alan P and Duyn, Jeff H},
  journal={IEEE transactions on medical imaging},
  volume={28},
  number={4},
  pages={551--554},
  year={2008},
  publisher={IEEE}
}

@article{merkle2011transmit,
  title={Transmit B1-field correction at 7T using actively tuned coupled inner elements},
  author={Merkle, Hellmut and Murphy-Boesch, Joseph and Gelderen, Peter van and Wang, Shumin and Li, Tie-Qiang and Koretsky, Alan P and Duyn, Josef H},
  journal={Magnetic Resonance in Medicine},
  volume={66},
  number={3},
  pages={901--910},
  year={2011},
  publisher={Wiley Online Library}
}

@article{zanovello2022cosimpy,
  title={CoSimPy: An open-source python library for MRI radiofrequency Coil EM/Circuit Cosimulation},
  author={Zanovello, Umberto and Seifert, Frank and Bottauscio, Oriano and Winter, Lukas and Zilberti, Luca and Ittermann, Bernd},
  journal={Computer Methods and Programs in Biomedicine},
  volume={216},
  pages={106684},
  year={2022},
  publisher={Elsevier}
}

@article{beqiri2015comparison,
  title={Comparison between simulated decoupling regimes for specific absorption rate prediction in parallel transmit MRI},
  author={Beqiri, Arian and Hand, Jeffrey W and Hajnal, Joseph V and Malik, Shaihan J},
  journal={Magnetic resonance in medicine},
  volume={74},
  number={5},
  pages={1423--1434},
  year={2015},
  publisher={Wiley Online Library}
}

@inproceedings{zhang2009field,
  title={Field and S-parameter simulation of arbitrary antenna structure with variable lumped elements},
  author={Zhang, Rongxing and Xing, Yao and Nistler, Juergen and Wang, Jianmin},
  booktitle={Proc. ISMRM},
  volume={17},
  pages={3040},
  year={2009}
}

@article{kozlov2009fast,
  title={Fast MRI coil analysis based on 3-D electromagnetic and RF circuit co-simulation},
  author={Kozlov, Mikhail and Turner, Robert},
  journal={Journal of magnetic resonance},
  volume={200},
  number={1},
  pages={147--152},
  year={2009},
  publisher={Elsevier}
}

@article{yan2014optimization,
  title={Optimization of an 8-channel loop-array coil for a 7 T MRI system with the guidance of a co-simulation approach},
  author={Yan, Xinqiang and Ma, ChuangXin and Shi, Lei and Zhuo, Yan and Zhou, Xiaohong Joe and Wei, Long and Xue, Rong},
  journal={Applied Magnetic Resonance},
  volume={45},
  number={5},
  pages={437--449},
  year={2014},
  publisher={Springer}
}

@article{lu2025magnetic,
  title={Magnetic field probe-based co-simulation method for irregular volume-type inductively coupled wireless MRI radiofrequency coils},
  author={Lu, Ming and Liang, Hao and Zhu, Haoqin and Yan, Xinqiang},
  journal={Magnetic Resonance Imaging},
  volume={117},
  pages={110330},
  year={2025},
  publisher={Elsevier}
}

@article{kazemivalipour2025simulated,
  title={Simulated radiation levels and patterns of MRI without a Faraday shielded room},
  author={Kazemivalipour, Ehsan and Guerin, Bastien and Wald, Lawrence L},
  journal={Magnetic Resonance in Medicine},
  volume={94},
  number={2},
  pages={835--851},
  year={2025},
  publisher={Wiley Online Library}
}

@article{li2021electromagnetic,
  title={Electromagnetic simulation of a 16-channel head transceiver at 7 T using circuit-spatial optimization},
  author={Li, Xin and Pan, Jullie W and Avdievich, Nikolai I and Hetherington, Hoby P and Rispoli, Joseph V},
  journal={Magnetic resonance in medicine},
  volume={85},
  number={6},
  pages={3463--3478},
  year={2021},
  publisher={Wiley Online Library}
}

@article{froncisz1986inductive,
  title={Inductive (flux linkage) coupling to local coils in magnetic resonance imaging and spectroscopy},
  author={Froncisz, Wojciech and Jesmanowicz, Andrzej and Hyde, James S},
  journal={Journal of Magnetic Resonance (1969)},
  volume={66},
  number={1},
  pages={135--143},
  year={1986},
  publisher={Elsevier}
}

@article{wright1991arrays,
  title={Arrays of mutually coupled receiver coils: theory and application},
  author={Wright, Steven M and Magin, Richard L and Kelton, James R},
  journal={Magnetic resonance in medicine},
  volume={17},
  number={1},
  pages={252--268},
  year={1991},
  publisher={Wiley Online Library}
}

@inproceedings{sahara2007development,
  title={Development of inductively coupled wireless radio frequency coil for magnetic resonance scanners},
  author={Sahara, Tomohiro and Hashimoto, Shigehiro and Tsutsui, Hiroshi and Mochizuki, Shuichi and Yamasaki, Kenichi and Kondo, Hideo and Yamada, Eiji and Akazawa, Kenzo and Okuyama, Kazuo and Inoue, Yuichi},
  booktitle={2007 Inaugural IEEE-IES Digital EcoSystems and Technologies Conference},
  pages={464--467},
  year={2007},
  organization={IEEE}
}

@inproceedings{zhu2012novel,
  title={A novel highly homogeneous wireless birdcage resonator coil},
  author={Zhu, Haoqin and Fallah-Rad, Mehran and Lang, Michael and Schellekens, Wayne and Champagne, Kirk and Petropoulos, Labros},
  booktitle={Proc. Intl. Soc. Mag. Reson. Med},
  volume={20},
  pages={2644--2644},
  year={2012}
}

@inproceedings{zhu2012novel1,
  title={A novel multichannel wireless receive phased array coil without integrated preamplifiers for high field MR imaging applications},
  author={Zhu, Haoqin and Fallah-Rad, Mehran and Lang, Michael and Schellekens, Wayne and Champagne, Kirk and Petropoulos, Labros},
  booktitle={Proc. Int. Soc. Mag. Reson. Med},
  volume={20},
  pages={2788},
  year={2012}
}

@article{zhu2013wireless,
  title={Wireless phased array endorectal coil for prostate imaging},
  author={Zhu, Haoqin and Fallah-Rad, Mehran and Petropoulos, Labros},
  journal={Proc. Intl. Soc. Mag. Reson. Med ISMRM},
  volume={2732},
  year={2013}
}

@article{zhang2017sensitivity,
  title={Sensitivity enhancement of traveling wave MRI using free local resonators: an experimental demonstration},
  author={Zhang, Xiaoliang},
  journal={Quantitative Imaging in Medicine and Surgery},
  volume={7},
  number={2},
  pages={170},
  year={2017}
}

@article{yan2017improved,
  title={Improved traveling-wave efficiency in 7 T human MRI using passive local loop and dipole arrays},
  author={Yan, Xinqiang and Zhang, Xiaoliang and Gore, John C and Grissom, William A},
  journal={Magnetic resonance imaging},
  volume={39},
  pages={103--109},
  year={2017},
  publisher={Elsevier}
}

@article{alipour2021improvement,
  title={Improvement of magnetic resonance imaging using a wireless radiofrequency resonator array},
  author={Alipour, Akbar and Seifert, Alan C and Delman, Bradley N and Robson, Philip M and Shrivastava, Raj and Hof, Patrick R and Adriany, Gregor and Fayad, Zahi A and Balchandani, Priti},
  journal={Scientific reports},
  volume={11},
  number={1},
  pages={23034},
  year={2021},
  publisher={Nature Publishing Group UK London}
}

@article{mo2025near,
  title={A Near-Field Coupling Array Enables Parallel Imaging and SNR Gain in MRI},
  author={Mo, Zhiguang and Che, Shao and Du, Feng and Xiao, Enhua and Chen, Qiaoyan and Li, Nan and Jia, Sen and Tie, Changjun and Wu, Bing and Zhang, Xiaoliang and others},
  journal={Advanced Science},
  pages={e03481},
  year={2025},
  publisher={Wiley Online Library}
}

@article{wu2024wireless,
  title={Wireless, customizable coaxially shielded coils for magnetic resonance imaging},
  author={Wu, Ke and Zhu, Xia and Anderson, Stephan W and Zhang, Xin},
  journal={Science Advances},
  volume={10},
  number={24},
  pages={eadn5195},
  year={2024},
  publisher={American Association for the Advancement of Science}
}

@article{wang2008inducitve,
  title={Inductive Coupled Local TX Coil Design},
  author={W. Wang and X. Lu and J. You and W. Zhang and H. Wang and H. Greim and M. Vester and and J. Wang},
  journal={Proc. Intl. Soc. Mag. Reson. Med ISMRM},
  volume={1510},
  year={2008}
}

@article{zhu2024detunable,
  title={Detunable wireless Litzcage coil for human head MRI at 1.5 T},
  author={Zhu, Haoqin and Lang, Michael L and Yang, Yijin and Martin, Melanie and Zhang, Gong and Zhang, Qiang and Chen, Yuanyuan and Yan, Xinqiang},
  journal={NMR in Biomedicine},
  volume={37},
  number={3},
  pages={e5068},
  year={2024},
  publisher={Wiley Online Library}
}

@article{zhu2024detunable1,
  title={Detunable wireless resonator arrays for TMJ MRI: A comparative study},
  author={Zhu, Haoqin and Zhang, Qiang and Li, Rangsong and Chen, Yuanyuan and Zhang, Gong and Wang, Ruilin and Lu, Ming and Yan, Xinqiang},
  journal={Magnetic Resonance Imaging},
  volume={111},
  pages={84--89},
  year={2024},
  publisher={Elsevier}
}

@article{zhu2024detunable2,
  title={A detunable wireless resonator insert for high-resolution TMJ MRI at 1.5 T},
  author={Zhu, Haoqin and Zhang, Qiang and Li, Rangsong and Chen, Yuanyuan and Zhang, Gong and Wang, Ruilin and Lu, Ming and Yan, Xinqiang},
  journal={Journal of Magnetic Resonance},
  volume={360},
  pages={107650},
  year={2024},
  publisher={Elsevier}
}

@book{collins2016electromagnetics,
  title={Electromagnetics in magnetic resonance imaging: physical principles, related applications, and ongoing developments},
  author={Collins, Christopher M},
  year={2016},
  publisher={Morgan \& Claypool Publishers}
}

@article{lu2023investigating,
  title={Investigating Local Receive Arrays in tcMRgFUS System and Their Influence by Passive Antennas: A Simulation Study},
  author={Lu, Ming and Yan, Xinqiang},
  journal={IEEE Access},
  volume={11},
  pages={143998--144005},
  year={2023},
  publisher={IEEE}
}

@article{yan2023dark,
  title={Dark band artifact in transcranial MR-guided focused ultrasound: Mechanism and mitigation with passive crossed wire antennas},
  author={Yan, Xinqiang and Allen, Steven and Lu, Ming and Moore, David and Meyer, Craig H and Grissom, William A},
  journal={Magnetic resonance imaging},
  volume={103},
  pages={169--178},
  year={2023},
  publisher={Elsevier}
}

@article{shajan201416,
  title={A 16-channel dual-row transmit array in combination with a 31-element receive array for human brain imaging at 9.4 T},
  author={Shajan, Gunamony and Kozlov, Mikhail and Hoffmann, Jens and Turner, Robert and Scheffler, Klaus and Pohmann, Rolf},
  journal={Magnetic resonance in medicine},
  volume={71},
  number={2},
  pages={870--879},
  year={2014},
  publisher={Wiley Online Library}
}

@article{golestanirad2017feasibility,
  title={Feasibility of using linearly polarized rotating birdcage transmitters and close-fitting receive arrays in MRI to reduce SAR in the vicinity of deep brain simulation implants},
  author={Golestanirad, Laleh and Keil, Boris and Angelone, Leonardo M and Bonmassar, Giorgio and Mareyam, Azma and Wald, Lawrence L},
  journal={Magnetic resonance in medicine},
  volume={77},
  number={4},
  pages={1701--1712},
  year={2017},
  publisher={Wiley Online Library}
}

@article{yan2016simulation,
  title={Simulation verification of SNR and parallel imaging improvements by ICE-decoupled loop array in MRI},
  author={Yan, Xinqiang and Cao, Zhipeng and Zhang, Xiaoliang},
  journal={Applied magnetic resonance},
  volume={47},
  number={4},
  pages={395--403},
  year={2016},
  publisher={Springer}
}

@article{zhang2021effect,
  title={Effect of radiofrequency shield diameter on signal-to-noise ratio at ultra-high field MRI},
  author={Zhang, Bei and Adriany, Gregor and Delabarre, Lance and Radder, Jerahmie and Lagore, Russell and Rutt, Brian and Yang, Qing X and Ugurbil, Kamil and Lattanzi, Riccardo},
  journal={Magnetic resonance in medicine},
  volume={85},
  number={6},
  pages={3522--3530},
  year={2021},
  publisher={Wiley Online Library}
}

@article{lemdiasov2011numerical,
  title={A numerical postprocessing procedure for analyzing radio frequency MRI coils},
  author={Lemdiasov, Rostislav A and Obi, Aghogho A and Ludwig, Reinhold},
  journal={Concepts in Magnetic Resonance Part A},
  volume={38},
  number={4},
  pages={133--147},
  year={2011},
  publisher={Wiley Online Library}
}

@article{brui2022volumetric,
  title={Volumetric wireless coil for wrist MRI at 1.5 T as a practical alternative to Tx/Rx extremity coil: a comparative study},
  author={Brui, Ekaterina and Mikhailovskaya, Anna and Solomakha, Georgiy and Efimtcev, Alexander and Andreychenko, Anna and Shchelokova, Alena},
  journal={Journal of Magnetic Resonance},
  volume={339},
  pages={107209},
  year={2022},
  publisher={Elsevier}
}

@article{zhu2023helmholtz,
  title={Helmholtz coil-inspired volumetric wireless resonator for magnetic resonance imaging},
  author={Zhu, Xia and Wu, Ke and Anderson, Stephan W and Zhang, Xin},
  journal={Advanced Materials Technologies},
  volume={8},
  number={22},
  pages={2301053},
  year={2023},
  publisher={Wiley Online Library}
}

@article{jordan2019wireless,
  title={Wireless resonant circuits printed using aerosol jet deposition for MRI catheter tracking},
  author={Jordan, Caroline D and Thorne, Bradford RH and Wadhwa, Arjun and Losey, Aaron D and Ozhinsky, Eugene and Kondapavulur, Sravani and Fratello, Vincent and Moore, Teri and Stillson, Carol and Yee, Colin and others},
  journal={IEEE Transactions on Biomedical Engineering},
  volume={67},
  number={3},
  pages={876--882},
  year={2019},
  publisher={IEEE}
}

\section{Appendix}

\subsection*{Appendix I: Derivation of \( S_{\text{real\_ports}} \)}

The derivation of \( S_{\text{real\_ports}} \) aims to establish the relationship between \( b_{\text{real\_ports}} \) and \( a_{\text{real\_ports}} \).

By expanding the second row of Eq.~\eqref{Stransf_def}, we can get:
\begin{equation} \label{Srealport_1}
b_{\text{lumped\_ports}} = S_{\text{all\_ports},21} \, a_{\text{real\_ports}} + S_{\text{all\_ports},22} \, a_{\text{lumped\_ports}}
\end{equation}

Since \( a_{\text{lumped\_ports}} = S_{\text{LC}} \, b_{\text{lumped\_ports}} \), we substitute:
\begin{equation} \label{Srealport_2}
b_{\text{lumped\_ports}} = S_{\text{all\_ports},21} \, a_{\text{real\_ports}} + S_{\text{all\_ports},22} \, S_{\text{LC}} \, b_{\text{lumped\_ports}}
\end{equation}

Rearranging for \( b_{\text{lumped\_ports}} \):
\begin{equation} \label{Srealport_3}
(I - S_{\text{all\_ports},22} \, S_{\text{LC}}) \, b_{\text{lumped\_ports}} = S_{\text{all\_ports},21} \, a_{\text{real\_ports}}
\end{equation}

Solving for \( b_{\text{lumped\_ports}} \):
\begin{equation} \label{Srealport_4}
b_{\text{lumped\_ports}} = (I - S_{\text{all\_ports},22} \, S_{\text{LC}})^{-1} \, S_{\text{all\_ports},21} \, a_{\text{real\_ports}}
\end{equation}

Now, expanding the first row of Eq.~\eqref{Stransf_def}:
\begin{equation} \label{Srealport_5}
b_{\text{real\_ports}} = S_{\text{all\_ports},11} \, a_{\text{real\_ports}} + S_{\text{all\_ports},12} \, a_{\text{lumped\_ports}}
\end{equation}

Based on Eq.~\eqref{S_LC_Relation}, Eq.~\eqref{Srealport_4} and Eq.~\eqref{Srealport_5}, we can get:
\begin{align} \label{Srealport_6}
b_{\text{real\_ports}} 
&=  S_{\text{all\_ports},12}\;S_{\text{LC}}\; (I - S_{\text{all\_ports},22} \, S_{\text{LC}})^{-1} \, S_{\text{all\_ports},21} \, a_{\text{real\_ports}}\nonumber \\
&\quad + S_{\text{all\_ports},11} \, a_{\text{real\_ports}}\; \nonumber \\
&= \left[ S_{\text{all\_ports},12}\;S_{\text{LC}}\; (I - S_{\text{all\_ports},22} \, S_{\text{LC}})^{-1} \, S_{\text{all\_ports},21} \, + S_{\text{all\_ports},11} \right] \, a_{\text{real\_ports}}
\end{align}

Thus, the \( S_{\text{real\_ports}} \) is:
\begin{equation} \label{Srealport_7}
S_{\text{real\_ports}} =  S_{\text{all\_ports},12}\;S_{\text{LC}}\; (I - S_{\text{all\_ports},22} \, S_{\text{LC}})^{-1} \, S_{\text{all\_ports},21} \, + S_{\text{all\_ports},11}
\end{equation}

\par

\subsection*{Appendix II: Derivation of combined EM fields}
\par

The weights of fields aims to establish the relationship between \( a_{\text{real\_ports}} \) and \( a_{\text{lumped\_ports}} \):

Based on Eq.~\eqref{S_LC_Relation} and Eq.~\eqref{Srealport_4}, we can get:

\begin{equation} \label{Srealport_8}
S_{\text{LC}}^{-1}\; a_{\text{lumped\_ports}} =  (I - S_{\text{all\_ports},22} \, S_{\text{LC}})^{-1} \, S_{\text{all\_ports},21} \, a_{\text{real\_ports}}
\end{equation}

Solving for \( a_{\text{lumped\_ports}} \):

\begin{equation} \label{Srealport_9}
 a_{\text{lumped\_ports}} = S_{\text{LC}}\; (I - S_{\text{all\_ports},22} \, S_{\text{LC}})^{-1} \, S_{\text{all\_ports},21} \, a_{\text{real\_ports}}
\end{equation}

thus, the field is:

\begin{equation}
B_{\text{combined}} = \sum_{k=1}^{n_{\text{rp}}+n_{\text{lp}}} \text{B}_k \cdot \text{a}_k 
\quad \text{and} \quad
E_{\text{combined}} = \sum_{k=1}^{n_{\text{rp}}+n_{\text{lp}}} \text{E}_k \cdot \text{a}_k .
\end{equation}

\end{document}